\documentclass[apj]{emulateapj} 
\usepackage{natbib}
\usepackage{graphicx}

\slugcomment{Received ........; Accepted .........}

\shorttitle{Evidence for convection in Sunspot penumbra}

\shortauthors{Bharti et al.}

\begin{document}

\title{Evidence for convection in Sunspot penumbrae}

\author{L. Bharti$^{1}$, S. K. Solanki$^{1,2}$ and J. Hirzberger$^{1}$}
\affil{1. Max-Planck-Institute f\"ur sonnensystemforschung, Max-Planck-Str. 2, 37191 Katlenburg-Lindau, Germany}
\affil{2. School of Space Research, Kyung Hee University, Yongin, Gyeonggi Do, 446-701, Korea}
       \email{bharti@mps.mpg.de}

\begin{abstract}
We present an analysis of twisting motions in penumbral filaments
     in sunspots located at heliocentric angles from $30^\circ$ to
     $48^\circ$ using three time series of blue continuum images obtained by the
     Broadband Filter Imager (BFI) onboard {\it Hinode}. The relations of the
     twisting motions to the filament brightness and the position within the
     filament and within the penumbra, respectively, are investigated. Only certain portions of the filaments show twisting
     motions. In a statistical sense, the
     part of the twisting portion of a filament located closest to the umbra is brightest and possesses the fastest twisting
     motion, with a mean twisting velocity of 2.1\,km\,s$^{-1}$. The middle
     and outer sections of the twisting portion of the filament (lying increasingly further from the umbra), which are less bright, have mean
     velocities of 1.7\,km\,s$^{-1}$ and 1.35\,km\,s$^{-1}$, respectively. The
     observed reduction of brightness and twisting velocity towards the
     outer section of the filaments may be due to reducing upflow along the
     filament's long axis. No significant variation of twisting velocity
     as a function of viewing angles was found. The obtained correlation of brightness and velocity
     suggests that overturning convection causes the twisting motions observed in penumbral filament and may be
     the source of the energy needed to maintain the brightness of the filaments.

\end{abstract}

\keywords{Sun: convection -- sunspots -- Sun:
  granulation}

\section{Introduction}

A central question related to sunspots is how sufficient energy can be
transported through a strong magnetic field, that reaches well below the solar
surface (a deep penumbra, Solanki \& Schmidt 1993), to keep the penumbra as
bright as it is. It is generally assumed that  some form of magnetoconvection
exists in penumbrae, acting in conjunction with the complex structuring of the magnetic
and flow fields (see Solanki 2003 and Scharmer 2009 for  reviews). Recently, 3D radiative
MHD simulations have provided support for this conjecture (Heinemann et al.
2007; Rempel et al. 2009a; 2009b). Observational support has been less clear-cut, however.

      The twisting motion of penumbral filaments observed by Ichimoto et al.
      (2007) opens a new window to understanding the nature of penumbral filaments
      and potentially of small-scale magnetoconvection and penumbral energy transport.
      In sunspots located away from disk center, filaments oriented roughly
      perpendicular to the disk radius vector display a twisting motion that is always directed from
      the limb-side to the center-side of the filament. As deduced by Zakharov et al. (2008), these
      twisting motions are a manifestation of overturning convection,
      associated with bright filaments. According to Spruit et al. (2010) the dark stripes follow the direction of inclined magnetic field lines.
      In the present
      article we investigate these twisting motions in a large ensemble of
      filaments  and in different parts of filaments, allowing us to
      statistically determine their properties and the relationship with other
      parameters of the filaments. In particular, we are interested in
      determining whether the heat transport from below associated with
      these motions contributes to the observed brightness of the penumbra.

\section{Observations}

For the present study we selected time series of blue continuum (4500\,\AA ) images
of sunspots located away from disk center observed with the Broadband Filter Imager (BFI) of the Solar Optical Telescope (SOT)
onboard {\it Hinode} (Tsuneta et al. 2007). The first time series, recorded on
January 9, 2007, consists of 498 images at 30\,s cadence and contains a
sunspot (NOAA 10933) located at a heliocentric angle of $\theta = 48^\circ$. The other
two time series, covering another sunspot (NOAA 10923), were recorded on November 11
and 12, 2006, at heliocentric angles of $\theta = 38^\circ$ and $\theta =
30^\circ$, respectively. They contain 292 and 191 images, respectively, at a cadence of 20 s. The spatial
resolution in the blue continuum is 0\farcs2. The image scale is
0\farcs054 per pixel. Raw data were corrected for flat field and dark
current using the Solar-Soft pipelines for the Hinode SOT/BFI. A Wiener filter
(Sobotka et al. 1993) was applied to all images to correct the point spread
function of the telescope assuming diffraction on an ideal circular 50 cm aperture.
Finally, contributions of five-minute oscillations were filtered by means of a subsonic
filter (Title et al. 1989) with a cut-off velocity of 6\,km\,s$^{-1}$. The first and last 15 images from time series of January 9, 2007
and similarly 10 images from November 11 and 12, 2006 have been omitted due to the apodizing window used in the subsonic filtering. Intensities
in all images are normalized to the mean intensity of quiet regions close to the observed sunspot.
Time series of these observations have been presented by Ichimoto et al. (2007).

\begin{figure}
\vspace{4mm}
\hspace{35mm}
\centering

\includegraphics[width=70mm,angle=0]{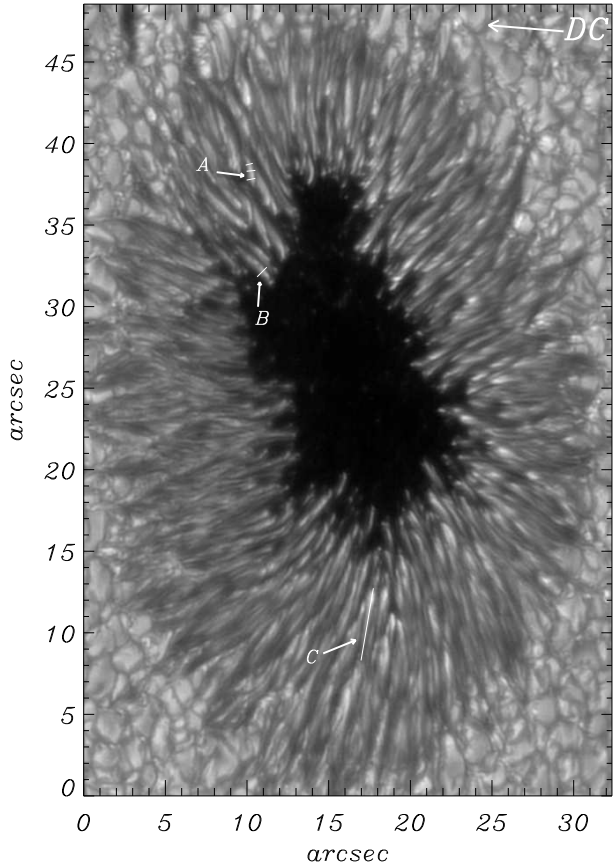}
\vspace{8mm}

\caption{A sunspot image in the blue continuum taken on January 9, 2007 by the SOT/BFI aboard Hinode.
The direction to disk center is shown by the arrow indicating 'DC' at the upper right corner. 'A' points to the location of
time slices at the inner cut, middle cut and outer cut (from below) of a penumbral filament
(see the main text and Fig. 2 for definations). Location of time slices for another penumbral
filaments is marked by 'B'. The location of the intensity profile along filament 'C' is marked by a white line.}
\end{figure}

\section{Analysis}

Figure~1 shows a blue continuum image as obtained on January 9, 2007. The
sunspot is located at $\theta = 48^\circ$. An animation of the time series of
such blue continuum images reveals a lateral motion of dark lanes in the
filaments from the limb-side to the disk-center-side in penumbral sections where the filaments
are oriented at angles smaller than $\pm$$60^\circ$ to the nearest part of the solar limb.
We have studied such motions in relation to other properties of 21 filaments in the northern section and 22 in the southern section.
We note that such twists are discernable only within a certain
portion of the length of a filament. We divided each filament into 3
parts (cf. Fig. 2), a central part displaying a twist (henceforth called the body of the
filament) and two parts bounding this central part that do not display any discernable twisting motions.
These include the 'head' of a filament i.e. the bright innermost part of the filament, and the 'tail' of a
filament, the often faint outer part in which no twist is seen anymore in
 space-time diagrams. Here, 'inner' and
'outer' parts of filaments refer to the parts closer to the inner and to the
outer boundary of the penumbra, respectively. The locations of the boundaries between head, body and tail vary
      from filament to filament. E.g., we find that filaments which extend into
      the umbra, possess twist over their whole
      length, although in the darker outer (i.e. away from umbra) part of the filament it becomes difficult to identify such a twisting motion.
      We call the length of the twisting portion (body) of the filament its twisting length.

      The 'twisting length' defined above may not be constant over the time series for each filament.
      The twisting length of filaments discussed later (Fig. 4 (right)) refers to the beginning of the
      time series. The twisting length of filaments which extend into the umbra, displays some ambiguity.

\begin{figure}
\vspace{3mm}
\hspace{-7mm}
\includegraphics[width=95mm,angle=0]{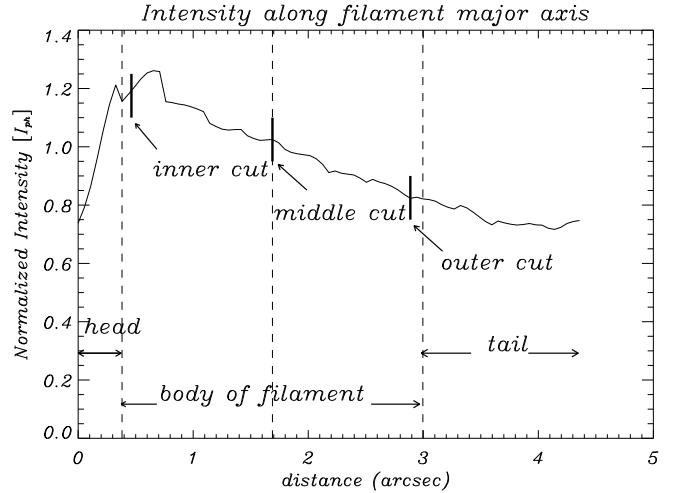}
\caption{Normalized intensity profile along the major axis of a typical penumbral filament. The plotted profile corresponds to the filament marked
'C' in Fig. 1. The intensity decreases from the bright head towards the outer section (i.e. with distance from umbra).
 Various terms defined in the main text are illustrated.}
\end{figure}

\begin{figure*}
\vspace{-100mm}
\hspace{5mm}
\includegraphics[width=1\textwidth,angle=0]{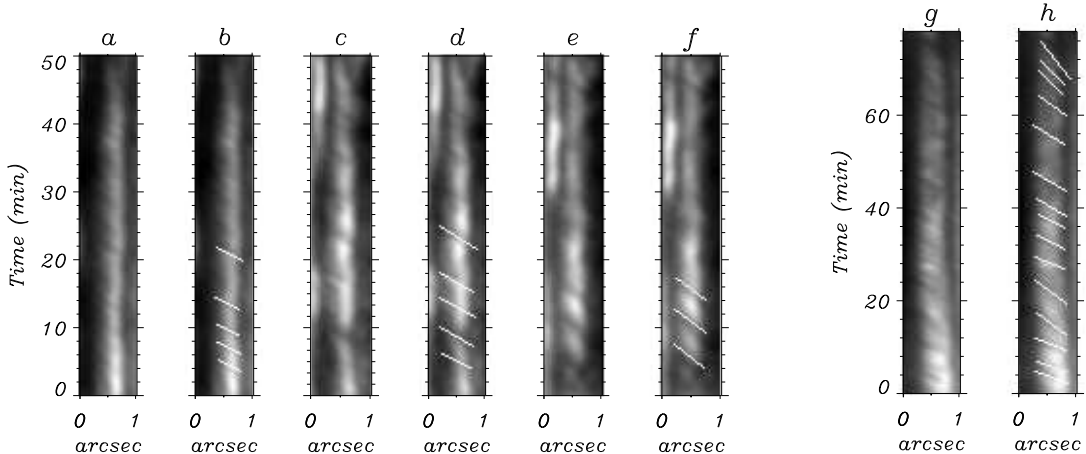}
\vspace{-105mm}
\caption{Space-time slices (a to f) at the three locations marked by 'A' in Fig. 1. Panels a and b refer to the inner cut, c and d to the
middle cut and e, f to the outer cut of filament 'A' (see Fig. 2 and the main text for a definition of these terms).
Panels b, d and f are the same as a, c and e, but with lines overlying the dark stripes to better reveal their slopes, which are
indicative of proper motion velocity. Panel g and h are space-time slices for filament 'B' at a fixed slice position. Panel h repeats panel g, but with white
lines overplotting the dark stripes.}
\end{figure*}

      We made space time diagrams for three locations in the body of the filament.
These three locations are the 'inner cut' which is close to the inner edge of the body of the filament, the 'middle cut' is the middle point of the
body of the filament and the 'outer cut' which is 2-3 pixels inside the outer edge of the body of the filament, such that dark stripes can still be seen
in the space-time diagram. The locations of all these parts
are sketched for filament 'C' in Fig.~2. The location of this particular filament 'C' is marked in
Figure~1. Note that the absence of a clear
signal does not imply the absence of corresponding motions in the head or
tail of a filament, just the absence of dark stripes parallel to the
filament's axis that make such motions visible.

The width of a (twisting) filament is the distance between the
inflection points of the intensity profiles parallel to the filament's minor axis.

      From Fig.~1 one can see that the inner part of a typical penumbral filament is brighter and the brightness
      decreases gradually towards the outer part. This is generally true, irrespective of whether the filament is located in the inner, middle or outer penumbra.
      The intensity profile along a filament's major axis as plotted in Fig. 2,
      is thus relatively typical of many filaments.

      Space-time diagrams of various locations, marked by arrows and white lines in Fig. 1, in some of the studied
filaments are shown in Fig. 3.
The motion of thin dark stripes across the filaments is apparent as inclined
dark stripes in Figs. 3, producing the impression of twisting
filaments,
 or, more precisely, of horizontal motions perpendicular to the filament's
major axes from the limb-side to the center-side. We stress that in our data set these motions
are manifested only by brightness variations and need not necessarily
correspond to a physical twist of the filaments. The slopes of these stripes
are proportional to the velocities of the horizontal motions of dark
structures in the filaments. The more horizontal the black stripe, the more
rapid the motion.

In the space-time diagrams displayed in Fig.~3 several successive twists can
be recognized. The slopes of the dark stripes are determined from pixels
 close to the left and right edges of the filament, respectively. The slopes of
two successive dark stripes are, with some scatter, almost the same if the brightness of the
filament does not change considerably.  The mean twist velocity is defined as the
mean velocity of two successive twists at the same location in the filament and the error of the velocity
measurement is assumed to be the difference between the two such velocity measurements. The mean peak brightness
is determined by averaging the maxima of the intensity profiles at several positions along the filament's minor axis between two successive twists.
 We have repeated this analysis at the inner, middle and outer cut, defined above. The time between two successive twists is found to be approximately 3 to 7 minutes. Within this
period the motion of bright penumbral grains along the major axes of the
filaments (cf. Sobotka et al. 1999) is small and leads only to negligible errors in the velocity of the twisting
motions.

      Figure~3 (a to f) shows time slices of filament 'A', at the inner, middle
and outer cuts from left to right (i.e. Fig. 3 a, c and e, respectively). Adjacent to each
image, the same image is displayed again, but with white lines overplotted on the dark stripes for better visibility. The
dependence of the twisting velocity (e.g. the slope of the dark stripes) on the
brightness of the body of the filament can be seen. The inner cut, which is the brightest, exhibits the highest
velocity. The middle and outer cuts, which are less bright, show lower
velocities.

      Depicted in Fig. 3g is the time-slice of filament 'B' close to the
      inner cut in the body of filament at the start of the time series. Figure 3h is the same as 3g, but with
       white lines  overplotted on the dark
      stripes. With time this filament penetrates increasingly into the umbra of
      the spot so that over time the fixed cutting position scans nearly the entire
      length of the filament. The variation of the twisting velocity with time, and
       hence with the location in the filament, is well illustrated.

      We find a variation of the velocity with location in the filament for nearly each of the 43 studied filaments of this spot.
      The same behavior is seen, irrespective of a filament's location in the inner, middle or outer penumbra.
       In general, the filament's intensity decreases over its body, from head to tail, as does the twisting velocity.

Twisting penumbral filaments are also found in the time series of the
spots located at $\theta = 38^\circ$ and $\theta = 30^\circ$,
respectively. In these time series in total 20 twisting filaments were
studied. No significant dependence of the twisting velocities on the
 viewing angle was found. The only difference to the spot located at
$\theta = 48^\circ$ is that in this latter spot the normalized intensities of the filaments are higher.

\begin{figure*}
\vspace{5mm}
\hspace{12mm}
\includegraphics[width=0.9\textwidth,angle=0]{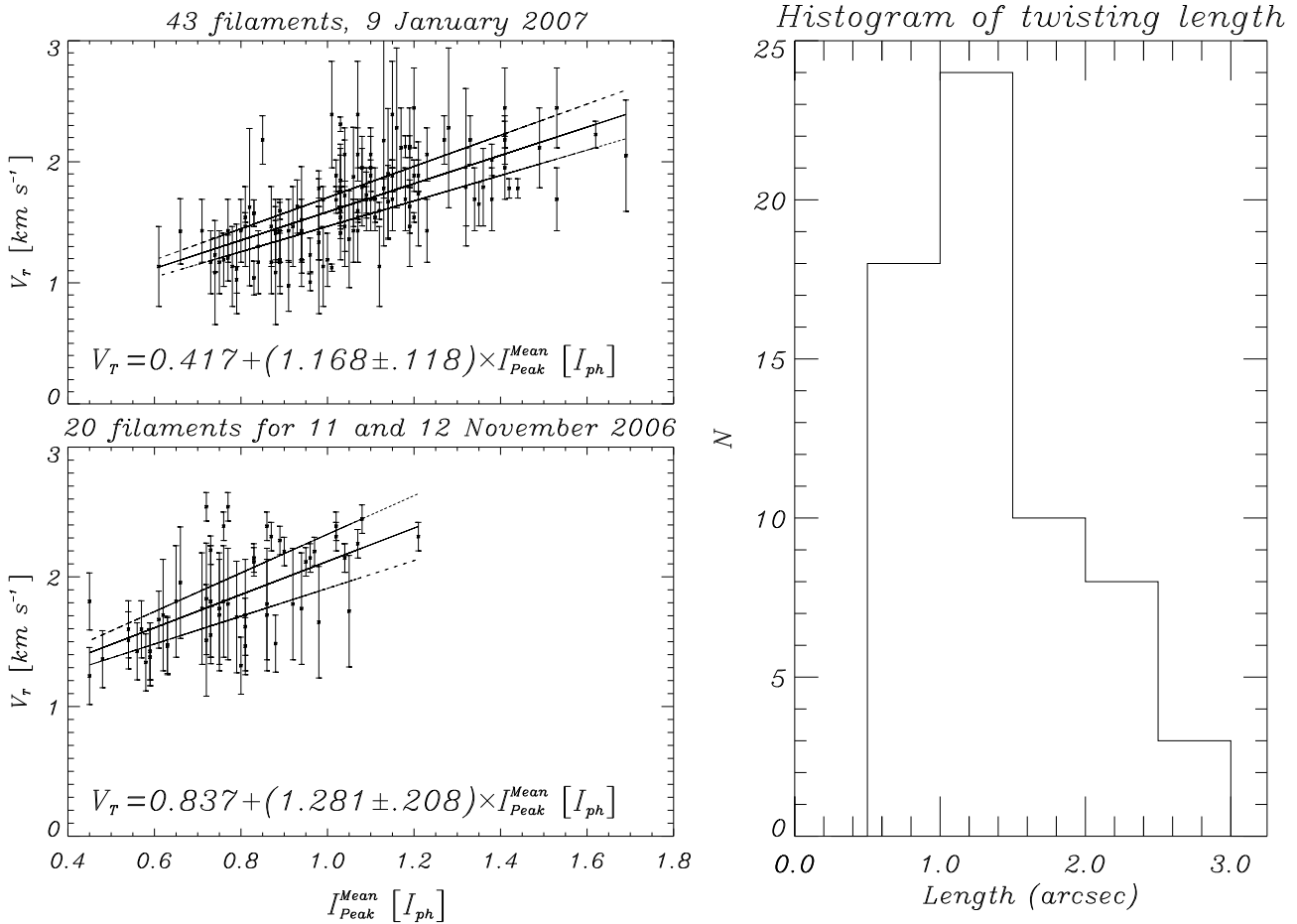}
\vspace{5mm}
\caption{Left: Scatter plots for mean peak intensity and twisting velocity for all cuts. Upper panel: scatter plot for 43 twisting filaments
observed on 9 January 2007 at $\theta = 48^\circ$. Lower panel: scatter plot for all 20 twisting filaments observed on
11 and 12 November 2006 at $\theta = 38^\circ$ and $\theta =
30^\circ$. The solid lines represent regression fits while the dashed lines
indicate $\pm$$\sigma$ uncertainty in the regression gradient. Right: Histogram of twisting lengths of filaments for all three time series.}
\end{figure*}

      In Fig.~4 (left) scatter plots of the twisting velocity vs. mean peak intensity
      at all three locations of the body of the filaments are shown. The upper panel
      displays the relationship for the 43 filaments of the spot
located at $\theta = 48^\circ$, for which a total of 129 data points are plotted (i.e. 3 cuts per filament). The error bars
represent the error of each velocity measurement, defined above. The solid line shows a linear fit
and the $\pm 1\sigma$ standard deviation of the fit from the data points is
indicated by dashed lines. Obviously, the "twisting" motion is stronger in
brighter parts of filaments. This is quantified by the linear correlation
coefficient, $C~=0.66$, with the probability of a chance correlation lying below 10$^{-4}$. The lower panel shows the
same for all 20 filaments extracted from the analyzed
sunspot at $\theta = 38^\circ$ and $\theta =
30^\circ$. For this plot the correlation coefficient is slightly reduced to
 $C~=0.63$, but the probability of a chance correlation remains below 10$^{-4}$.
In addition, the slopes of the linear regressions in both panels are significant at the 5-10$\sigma$ level (see the regression
 equations given in the upper and lower panels of Fig. 4 (left)). Consequently, both correlations are statistically significant.
 The lower normalized intensity
of the filaments of the spot located at $\theta = 38^\circ$ and $\theta =
30^\circ$ (lower panel) relative to the spot located at $\theta = 48^\circ$,
(upper panel) is evident.

In Fig.~4 (right) a histogram of the twisting length (body) of the filaments from all three
time series is depicted. The twisting lengths lie in the range between $0\farcs5$
and $3\farcs0$. The mean length of the twisting portion is $1\farcs4$. In
general, the head of the filament, which does not display any twisting motion, has a length about 10\,\% to 20\,\% of the entire filament.
However, filaments which extend into the umbra or which are located
in the inner penumbra show twists along their entire length, although this is
not always clear cut due to the increasing darkening and hence decreasing
visibility of stripes in the outer parts of the filaments. In most filaments twists can be
seen over fractions of roughly 50\,\% to 70\,\% of their entire length.

Histograms of filament width, mean peak intensity and twisting velocity for
inner, middle and outer cuts as obtained from all time series are
plotted in Fig.~5. The mean filament widths at the inner, middle and outer cuts are
$0\farcs27\pm0.08$, $0\farcs30\pm0.09$ and $0\farcs28\pm0.09$,
respectively. Consequently, the widths of the filaments are statistically similar over their twisting length (body of filament).
The mean peak intensities are found to be 1.14,
0.97 and 0.79 I$_{ph}$, respectively, for the three locations. The corresponding mean
twisting velocities are  2.11\,km\,s$^{-1}$,
1.67\,km\,s$^{-1}$ and 1.35\,km\,s$^{-1}$, respectively.

      Scatter plots (not shown here)
of width vs. intensity, width vs. velocity and twisting length vs. twisting velocity at inner, middle
 and outer cuts  reveal no significant correlations. However,
filaments having larger twisting lengths show lower brightness in middle and outer cuts.

\begin{figure*}
\vspace{5mm}
\centering
\includegraphics[width=.9\textwidth,angle=0]{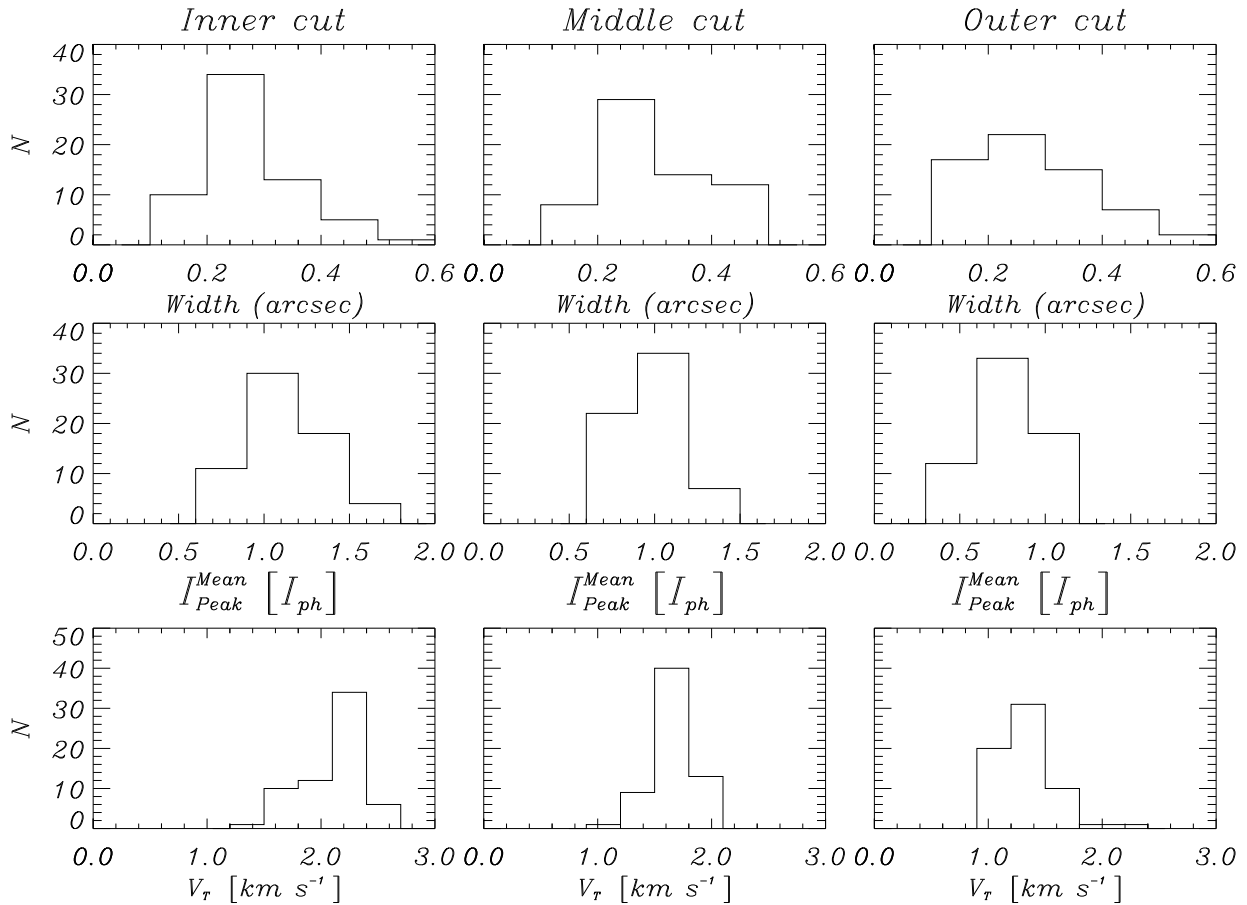}
\vspace{5mm}
\caption{Histograms of widths (top panels), normalized mean peak intensities (middle) and twisting velocities (bottom)
for inner, middle and outer cuts for all three time series.
From left-to right the histograms for inner, middle and outer cuts are plotted. }
\end{figure*}

\section{Discussion and conclusion}

We analyzed twisting motions at different positions in penumbral filaments viewed from different (heliocentric) angles as
observed in time series of seeing-free blue continuum images obtained with
{\it Hinode}. We find that the apparent speed of the twisting motion is related to the local brightness. The parts of the filaments closest to the umbra show the
highest brightness and fastest twisting motion.
The parts of the filaments further from the umbra are less bright and have lower twist velocities.

      These "twisting" motions are most naturally interpreted in terms of overturning convective flows
perpendicular to the filament's major axis (Zakharov et al. 2008, Scharmer 2009, Spruit et al. 2010). In particular, Zakharov et al. (2008)
have established that such convective motions can transport sufficient energy to explain the brightness of the
penumbra. The brightness variation along the filament's length suggests that such small-scale convection
(directed perpendicular to the axis of the filament) is more efficient around
the head of the filament and its efficiency decreases gradually towards the
filament's tail.

      The Evershed effect has been invoked multiple times as a
      source of heating of the penumbral gas. Schlichenmaier et al. (1998a, b) proposed
      hot gas to flow into the penumbral photosphere through the
      cross-section of flux-tubes (cf.
      Solanki and Montavon 1993). This turns out to be insufficient to heat
      the penumbra as a whole (Schlichenmaier and Solanki 2003). Scharmer et
      al. (2008a) and Rempel et al. (2009a, b) have proposed that the Evershed
      flow in a given filament forms a heavily elongated convective cell
      which is mainly responsible for keeping the penumbra bright. In this
      picture, filaments carrying the Evershed flow have a reduced, nearly horizontal magnetic field and
       gas rises and submerges over large
      parts of the filament, greatly enhancing the efficiency of convection.
      Importantly, the gas rising near the axis of symmetry of the filament flows down not just
      around the tail of a filament, but also at its sides. This component of the flow moving sideways across the filaments is
      the likely cause of the 'twisting' motions investigated here.

            The fact that this clearly convective component of the flow is
            correlated with the brightness of the filament harboring it
            significantly strengthens the evidence that the overturning
            convection is responsible for heating the penumbra and in
            particular supports the validity of the energy flux estimate
            made by Zakharov et al. (2008)\footnote[1]{A chance correlation cannot be completely ruled out,
            however, since rising and sinking of the Evershed flow in the inner and outer penumbra could
            also produce the observed brightness gradient.}. Also, it raises the question of
            the geometry of the flow and of the associated magnetic field in
            penumbral filaments, which is so far unresolved.  Note that
            observations clearly indicate the presence of a magnetic field in
            the Evershed flow channels (Solanki et al. 1994) of kG strength
            (Borrero et al. 2005; Borrero \& Solanki 2008, cf. Bellot Rubio
            et al. 2004, 2007, Jur\v c\'ak et al. 2007, Scharmer et al. 2008b).
            Due to the presence of flow components directed along as well as
            across the axis of the filament, such field lines get twisted.

      From the observations alone it is not possible to determine how deep
      the covective down and upflows reach below the surface. Recently,
      Spruit et al. (2010) have argued for deep convection. Their
      criticism of the roll-like convection interpretation is partly based on
      the misconception that these rolls need to have a circular
      cross-section. Of course, they can extend much deeper.
      However, the interpretation of convective rolls does imply that at
      some point below the surface the surrounding magnetic field wraps around the
      convective fluid again, separating it from the field-free convecting gas below the penumbra.
      This is in contrast to the field-free gap model of
      Spruit and Scharmer (2006), cf. Scharmer (2009) in which the field-free
      gap is open at the bottom. Numerical simulations (Heinemann et al. 2008,
      Rempel et al. 2009) indicate that the convective features reach deep, but are indeed
      wrapped by the surrounding prnumbral field and do not open into the
      field-free gas surrounding the sunspot. This is the main difference
      between the two pictures, besides the absence of a
      magnetic field within the gaps (Spruit and Scharmer 2006), but its
      presence in the rolls proposed by Zakharov et al. (2008).

                 The present paper has provided further support for the idea
                 that overturning convection is an important process in
                 maintaining the penumbral temperature. The clear signal of
                 an upflow along the central axis of a bright filament
                 has been reported by Franz \& Schlichenmaier (2010) and Ichimoto
                 (2010) but downflows at its sides has still to be seen, however,
                 although indirect evidence has been found by M\'arquez et al.
                 (2006) and S\'anchez Almeida et al. (2007).

\acknowledgments

This work has been partly supported by the WCU grant No. R31-10016 funded by
the Korean Ministry of Education, Science and Technology.
Hinode is a Japanese mission developed and launched by ISAS/JAXA,
collaborating with NAOJ as a domestic partner, NASA and STFC (UK) as
international partners. Scientific operation of the Hinode mission is
conducted by the Hinode science team organized at ISAS/JAXA.
Support for the post-launch operation is provided by JAXA
and NAOJ (Japan), STFC (U.K.), NASA (U.S.A.), ESA, and NSC (Norway).

\end{document}